\documentclass[times,twocolumn,final]{elsarticle}

\usepackage{cnf}

\usepackage{framed,multirow}

\usepackage{amssymb}
\usepackage{latexsym}
\usepackage{amsmath}

\usepackage{url}
\usepackage{xcolor}
\definecolor{newcolor}{rgb}{.8,.349,.1}

\usepackage{hyperref}
\usepackage{verbatim}

\usepackage[switch,pagewise]{lineno} 

\begin{document}

\verso{Kushwaha et al.}

\begin{frontmatter}

\title{Role of flame localization in wavemaker regions on the suppression of combustion instability in turbulent partially-premixed methane flames}%
\tnotetext[tnote1]{Abhishek Kushwaha}

\author[1]{Abhishek \snm{Kushwaha} \corref{cor1}}
\cortext[cor1]{Corresponding author: Abhishek Kushwaha}
\emailauthor{abhikushwaha8090@gmail.com}{Abhishek Kushwaha}
\author[1]{Bjarne \snm{Lieth}}    
\author[1]{Isaac \snm{Boxx}}

\address[1]{Lehrstuhl für Optische Messverfahren für die Energie- und Verfahrenstechnik (LOM), RWTH Aachen University, Germany.}

\begin{abstract}




The interaction between flame dynamics and hydrodynamic instabilities plays a fundamental role in determining the stability of swirl-stabilized combustors. In the present study, we investigate the hypothesis that the spatial overlap between the flame and the wavemaker region is a necessary prerequisite for combustion stability in flows characterized by a wavemaker region. To test the hypothesis, a low-momentum secondary methane injection was introduced through circumferential holes located on the centerbody of a swirl-stabilized burner. The injection velocity was maintained below 5\% of the bulk flow velocity to minimize momentum-induced modifications of the flow field while selectively redistributing the heat release. In addition, an equivalent amount of fuel was diverted from the primary fuel supply to the secondary injection ports, known as fuel-staging, to isolate the effects of flame relocation from those of the total fuel flow rate. The introduction of secondary injection produced a transition of the flame from M-shaped to V-shaped structure, while fuel-staging yielded a similar flame response, demonstrating that the observed behavior results primarily from the redistribution of heat release. Linear stability analysis revealed that the wavemaker region remains near the inlet of the combustion chamber. Moreover, the flame root transitions from a lifted M-flame to an attached V-flame and consistently stabilizes at the radial position corresponding to the identified wavemaker region. The coincidence of flame attachment and wavemaker location under stable operating conditions provides strong experimental evidence that flame stabilization is governed by their spatial overlap rather than by a modification of the underlying hydrodynamic instability. The proposed framework provides new insight into the coupling between flame stabilization and hydrodynamic instability and offers practical guidance for the design of stable, low-emission combustion systems.

\end{abstract}


\end{frontmatter}

\linenumbers

\section*{Novelty and significance statement}


The novelty of the work lies in experimentally validating the spatial overlap between the flame and the wavemaker region as a governing criterion for combustion stability in swirl-stabilized combustors. While the wavemaker concept is well established in hydrodynamic stability analysis for identifying the region governing global instability mode selection, its experimental link to flame stabilization in reacting flows has remained largely unexplored. 



The findings from this study provide direct experimental support for the hypothesis that spatial overlap between the flame and the wavemaker region is a key prerequisite for combustion stability, establishing flame–wavemaker overlap as a physically meaningful criterion for passive flame stabilization.


\section{Introduction}
\label{sec1}

Gas turbine companies need to comply with very tight environmental standards and international regulations about the emissions, which have led to the emergence of lean premixed swirl-stabilized combustion systems \cite{international2014sustainable}. Even though lean premixed combustion systems reduce NO$_x$ emissions \cite{lazik2008development}, these systems impose an important limitation such as an increased susceptibility to combustion instabilities \cite{correa1993review, lieuwen1998role, nakata1998reaction}. Combustion instability is a central problem in modern combustion research, arising when unsteady heat release couples with chamber acoustics \cite{poinsot1987vortex, mcmanus1993review,pawar2017thermoacoustic} and satisfies Rayleigh's criterion \cite{rayleigh1878explanation}. The coupling among the subsystems produces high-amplitude, self-sustained periodic oscillations  \cite{sujith2021thermoacoustic} that lead to structural fatigue, component damage, and, in severe cases, failure of engine hardware \cite{shanbhogue2009lean}. The acoustic signal shows the in-phase or constant phase relation with the heat release rate signal during combustion instability \cite{nicoud2005thermoacoustic}. In swirl-stabilized combustors, the reacting flow is governed by the interplay of shear layer dynamics \cite{candel2002combustion}, vortex breakdown\cite{lambourne1961bursting,escudier1982vortex}, and flame stabilization \cite{zhao2018experimental} processes, all of which collectively determine the overall stability of the system.

A key feature of swirling flows is the formation of an inner recirculation zone (IRZ) associated with vortex breakdown \cite{freitag2005mixing}. The formation of the inner recirculation zone is frequently, though not invariably, associated with coherent flow structures such as the precessing vortex core (PVC) \cite{freitag2005mixing,syred2006review,stohr2011dynamics}. Experimental observations and numerical simulations such as large eddy simulation (LES), have consistently demonstrated the presence of the PVC in both reacting \cite{selle2004compressible,schildmacher2006experimental,freitag2007investigation} and non-reacting \cite{patel2008simulation,stohr2012experimental} swirl flows. The PVC interacts strongly with the flame, modulating its shape and position. This interaction becomes particularly significant in the broader context of combustion instability, as PVC-induced flow oscillations can introduce periodic perturbations to the heat release, potentially contributing to or altering the combustor system's stability characteristics \cite{allison2013experimental}. The presence and strength of the PVC are highly sensitive to operating parameters, including swirl number \cite{duwig2007large}, thermal power \cite{steinberg2013parametric}, and inlet conditions \cite{terhaar2015vortex}. Notably, modifications to these parameters can suppress the PVC, often accompanied by significant changes in flame topology and, consequently, combustion dynamics.

Furthermore, the PVC induces periodic modulation of the velocity and scalar fields \cite{moeck2012nonlinear, rukes2016impact, balasubramaniyan2021global, karmarkar2022impact}, which in turn affects the local equivalence ratio, flame surface area, and heat release. This coupling mechanism can be particularly pronounced in partially premixed or stratified combustion regimes, where the sensitivity of the flame to flow perturbations is higher \cite{oberleithner2015formation}. The spatial structure of the PVC, with its helical deformation of the flow, introduces asymmetry in the flame dynamics, which enhance localized heat release fluctuations \cite{terhaar2015vortex}. Experimental studies have shown that suppression or alteration of the PVC, through geometric modifications or flow control strategies, can significantly affect the flame dynamics of the combustor, thereby influencing the combustion dynamics of the system \cite{duwig2007large}. 

Flame shape and topology play a crucial role in determining the combustion dynamics of swirl combustors, as these characteristics directly influence the spatial and temporal distribution of heat release \cite{stohr2009experimental,chterev2014flame}. In swirl-stabilized flames, common configurations include V-shaped, M-shaped, and conical flame structures \cite{kushwaha2021dynamical} and these flame shapes are governed by a complex interplay of multiple factors, including swirl intensity, inlet velocity, equivalence ratio, and combustor geometry. Previous studies have shown that certain flame shapes, such as M-shaped flame, are more susceptible to combustion instability, particularly when the flame configuration allows for strong feedback between flow perturbations and heat release \cite{oberleithner2015formation,kushwaha2024coupled}. The M-shaped flame that interacts with both inner and outer recirculation zones, exhibit enhanced sensitivity to vortex-induced perturbations. Moreover, the anchoring location of the flame relative to the recirculation zones and shear layers plays a critical role in defining its dynamic behavior \cite{huerre1985absolute}. The relationship between flame shape and combustion dynamics along with global instability also influence the flow field, specially to regions of maximum instability such as the wavemaker regions \cite{rukes2016impact}. 

To better understand the mechanisms governing global instability, the concept of the wavemaker region  provides a strong framework for identifying the spatial origin of such instability in fluid flows, including swirl combustion systems \cite{monkewitz1982influence, chomaz2005global,giannetti2007structural}. The wavemaker is defined as the region in the flow where perturbations are both most receptive and most influential in sustaining a global mode, effectively acting as the core of the instability mechanism \cite{chomaz2005global,huerre1990local}. In swirling flows exhibiting a precessing vortex core, the wavemaker region is often located near the vortex breakdown zone, where the flow transitions from axial to recirculating and where shear layers are strongest \cite{oberleithner2015formation,o2023understanding}. While this concept has been successfully applied in hydrodynamic stability analyses, its role in reacting flows, particularly in relation to flame stabilization, remains insufficiently understood. Specifically, the extent to which the spatial positioning of the flame relative to the wavemaker region influences combustion instability has not been conclusively established.

Recent studies suggest that localized regions of high structural sensitivity such as wavemaker region, play a critical role in instability development, as these structures facilitate strong feedback between flow perturbations and system response \cite{karmarkar2022impact, manoharan2020weakly, gupta2022impact}. However, direct experimental evidence linking these sensitive regions to the spatial distribution of heat release in practical combustors is still limited. The challenge is compounded by the fact that flame position is not fixed but depends on a range of factors, including inlet conditions, mixing processes, and geometric confinement. Even subtle variations in mixture composition can alter flame anchoring and modify its interaction with the underlying flow structures  \cite{chong2016effect}. In particular, localized fuel enrichment near the centerbody offers a potential means of influencing flame stabilization without significantly altering the global flow field. Such modifications can shift the heat-release distribution and potentially affect the density field in regions critical to instability development \cite{oztarlik2020suppression}. Despite this, the impact of weak, spatially confined fuel addition on the coupling between flame dynamics and wavemaker regions has not been systematically explored.

The present study tests the hypothesis that spatial overlap between the flame and the wavemaker region is a necessary prerequisite for combustion stability in swirl-stabilized combustors. 
For this purpose, controlled secondary methane injection was introduced through azimuthally distributed ports along the burner, maintaining an injection velocity well below 5\% of the bulk flow velocity. This ensures that the primary effect of the secondary fuel is the local redistribution of heat release rather than any significant alteration of the flow field's overall momentum. The approach enabled targeted relocation of the flame root while preserving the underlying flow configuration. The wavemaker location remained anchored near the combustor inlet throughout the experiments. Complementary fuel-staging tests, in which an equivalent amount of fuel was diverted from the primary to the secondary injection ports, confirmed that the flame response arose from spatial redistribution of heat release rather than from changes in the overall fuel flow rate. Under both secondary injection and fuel-staging, the flame transitioned from a lifted M-shaped structure to an attached V-shaped structure. The flame root consistently anchored within the radial band corresponding to the wavemaker region identified by linear stability analysis. Coincidence between flame attachment and wavemaker location, observed under stable operating conditions, provides direct experimental evidence that spatial overlap between flame and wavemaker region governs combustion stability. Flame–wavemaker overlap thus emerges as a physically meaningful and practically actionable framework for passive flame stabilization in swirl-stabilized combustors.

\section{Theoretical fundamentals}
\label{The_Fun}

\subsection{Location of wavemaker region in velocity field}
\label{location}

The wavemaker region is identified using a local linear stability analysis (LSA) performed on the ensemble-mean velocity field $\bar{\textbf{u}}(\textbf{x})$, obtained using Stereo-PIV measurements. The instantaneous flow velocity field is a vector with radial, axial, and azimuthal components, written as

\begin{gather} 
\mathbf{u}(x,y,t)
 =
   \begin{bmatrix}
   u_r(x,y,t_k) \\
   u_y(x,y,t_k) \\
   u_\theta(x,y,t_k) 
   \end{bmatrix}
   \label{eq1}
\end{gather}

where $u_r$, $u_y$ and $u_\theta$ are the radial, axial and azimuthal components of the velocity fields. $N_{PIV}$ represents the total number of velocity fields acquired, i.e. 400. 

Its time-averaged part $\bar{\mathbf{u}}(\mathbf{x})$ is obtained by ensemble averaging over $N_{PIV}$ realizations as,

\begin{equation} 
    \overline{\mathbf{u}}(x,y) = \frac{1}{N_{PIV}}
\sum_{k=1}^{N_{PIV}}
\mathbf{u}(x,y,t_k).
\label{eq2}
\end{equation}

The fluctuating part is represented as

\begin{equation}
    {\mathbf{u}^{\prime}}(\mathbf{x},t) = 
    \mathbf{u}(\mathbf{x},t) - \mathbf{\bar{u}}(\mathbf{x})
    \label{eq3}
\end{equation} 

Since the dominant hydrodynamic instability in swirling combustors is primarily governed by the axial and azimuthal velocity distributions, radial profiles of $(\overline{u}_r)$ and $(\overline{u}_{\theta})$ were extracted at each axial location. The burner centerline was determined at the geometric center of the measurement domain, i.e. $x_0 = 0$. For each axial station $y_j$, the radial profiles of the time-averaged axial velocity $\bar{u}_y(r)$ and azimuthal velocity $\bar{u}_{\theta}$ are extracted and sorted monotonically in $r$. Duplicate and non-unique radial locations are removed to ensure a well-posed discrete problem. The azimuthal velocity is enforced to vanish at the axis of symmetry ($\bar{u}_{\theta} = 0$ at $r = 0$) as required by the regularity conditions of the cylindrical coordinate system.

The local eigenvalue problem is formulated for the azimuthal mode $m = 1$, corresponding to the single-helical precessing vortex core (PVC) instability. Assuming a quasi-parallel base flow, the fluctuating field $\mathbf{u}^{\prime}(\mathbf{x},t)$ is expressed in normal-mode form as [Eq. \ref{eq4}], where $\hat{\mathbf{u}}(r)$ is the complex mode shape, $\alpha \in \mathbb{C}$ is the complex streamwise wavenumber and $\omega \in \mathbb{C}$ is the complex frequency.

\begin{equation} 
    \tilde{\mathbf{u}}(r,t) = \hat{\mathbf{u}}(r)\,
        e^{i(\alpha y - \omega t)} + \text{c.c.}
    \label{eq4}
\end{equation}

To efficiently scan the streamwise domain and locate the axial wavemaker station, a simplified local stability operator is first solved independently at each axial station $y_j$, retaining only the axial advection and viscous diffusion terms and omitting the swirl and cylindrical curvature contributions present in the full $m = 1$ formulation. The governing operator is discretized on $N_{PIV}$ radial grid points using the second-derivative finite-difference operator

The linearized governing operator is discretized on the
$N_{PIV}$ radial grid points using second-order finite differences. The discrete First-derivative operator ($\mathbf{D}_1$) and second-derivative operator ($\mathbf{D}_2$) in cylindrical geometry are constructed as

\begin{equation} 
\mathbf{D}_1 = \frac{1}{2\Delta r}\,\mathrm{tridiag}(-1,\,0,\,1),
\label{eq8}
\end{equation}

\begin{equation} 
\mathbf{D}_2 = \frac{1}{\Delta r^2}\,\mathrm{tridiag}(1,\,-2,\,1),
\label{eq5}
\end{equation}

with the grid spacing $\Delta r = \langle \mathrm{d}r \rangle$
computed as the arithmetic mean of consecutive radial increments.
The local stability operator is then assembled as

\begin{equation} 
    \mathbf{L} = -i\alpha\,\{ \mathrm{diag}(\bar{u}_y)\}
        + \nu_{\mathrm{eff}}\!\left(
            \mathbf{D}_2 - \alpha^2\mathbf{I}
          \right),
    \label{eq6}
\end{equation}

where $\nu_{\mathrm{eff}}$ denotes the effective kinematic viscosity. For the initial pass over all axial stations, $\alpha$ is prescribed as a complex initial guess ($\alpha = 1 + 0.1i$) and the eigenvalue problem $\mathbf{L}\hat{u} = \omega\hat{u}$ is solved by full-matrix eigendecomposition. A homogeneous Dirichlet condition is imposed at the innermost radial node. The eigenvalue $\omega_0(y_j)$ exhibiting the largest imaginary part (i.e.\ maximum temporal growth rate) is retained at each axial station.

At the axial location identified as the global wavemaker, a complete cylindrical stability operator is assembled to resolve the radial structure of the instability mode. The radial velocity profiles are first smoothed using locally weighted scatterplot smoothing with a span of 5\% of the data range to suppress measurement noise, and radial gradients are computed using central differences in the interior of the domain (with one-sided differences applied at the two boundary points):


\begin{equation}
    \begin{split}
     \frac{\mathrm{d}\bar{u}_y}{\mathrm{d}r}\bigg|_i \approx 
    \frac{\bar{u}_y(r_{i+1}) - \bar{u}_y(r_{i-1})}{r_{i+1}-r_{i-1}}, \\
    \frac{\mathrm{d}\bar{u}_{\theta}}{\mathrm{d}r}\bigg|_i \approx 
    \frac{\bar{u}_{\theta}(r_{i+1}) - \bar{u}_{\theta}(r_{i-1})}{r_{i+1}-r_{i-1}}
    \end{split}
    \label{eq7}
\end{equation}

The cylindrical Laplacian operator $\boldsymbol{\mathcal{L}}$ for azimuthal mode $m$ is described as

\begin{equation} 
    \boldsymbol{\mathcal{L}} = \mathbf{D}_2
        + \mathrm{diag}\!\left(\frac{1}{r}\right)\mathbf{D}_1
        - \mathrm{diag}\!\left(\frac{m^2}{r^2}\right)\mathbf{I}
        - \alpha^2\mathbf{I},
    \label{eq9}
\end{equation}

where the radial coordinate is regularized near the axis as $r \to r + \varepsilon$ with $\varepsilon = \Delta r/10$ to avoid division by zero. The convective term $ \mathbf{C}$ accounting for both axial and swirl base-flow contributions is

\begin{equation}
    \mathbf{C} = -i\!\left[
        \alpha\,\mathrm{diag}(\bar{u}_y)
        + m\,\mathrm{diag}\!\left(\frac{\bar{u}_{\theta}}{r}\right)
    \right],
    \label{eq10}
\end{equation}

yielding the full local operator

\begin{equation}
   \boldsymbol{\mathcal{Q}} = \mathbf{C} + \nu_{\mathrm{eff}}\,\boldsymbol{\mathcal{L}}.
    \label{eq11}
\end{equation}

The eigenvalue problem $\boldsymbol{\mathcal{Q}}\hat{u}(r) = \omega_0 \hat{u}(r)$ where ${\boldsymbol{\mathcal{Q}}}$ combines the linearized convective terms with the viscous cylindrical Laplacian operator, is solved to obtain the complex frequency $\omega_0$ and radial eigenfunction $\hat{u}(r)$ of each admissible disturbance mode at the wavemaker station.


The concept of the global mode wavemaker, following Chomaz \textit{et. al.} \cite{chomaz2005global} and Huerre \textit{et. al.} \cite{huerre1990local}, identifies the streamwise location $y_s$ at which the global oscillation frequency and growth rate are set. For a region of absolute instability attached to the flow inlet, the global mode frequency ($\omega_g$) equals the absolute frequency at the inlet \citep{oberleithner2015formation}:

\begin{equation} 
    \omega_g = \omega_0(y = 0), \qquad y_s = 0.
    \label{eq12}
\end{equation}

More generally, the wavemaker is located where the streamwise gradient of the absolute frequency $\omega_0$ vanishes:

\begin{equation}
    \left.\frac{\mathrm{d}\omega_0}{\mathrm{d}y}\right|_{y=y_s} = 0,
    \label{eq13}
\end{equation}

which corresponds to the saddle-point criterion of the global mode instability \citep{chomaz2005global}. In practice, the discrete gradient $\mathrm{d}\omega_{0,r}/\mathrm{d}y$ is evaluated using a second-order centered scheme, and the axial wavemaker location $y_s$ is identified as the streamwise station at which $|\mathrm{d}\omega_{0,r}/\mathrm{d}y|$ is minimized. The non-dimensionalized axial wavemaker coordinate is given by $y_s/D$, where $D = 27.85$\,mm is the inlet diameter of the combustion chamber.

To identify the radial extent of the wavemaker region from experimental data, we adapt the classical structural sensitivity concept of Giannetti \& Luchini \cite{giannetti2007structural}, in which the wavemaker is identified as the spatial overlap of the direct instability mode and its adjoint, i.e. $|u|\cdot |u^+|$, representing the region where the mode has significant amplitude and where the flow is maximally receptive to perturbations. Since the adjoint field cannot be obtained from experimental velocity data without a full numerical stability solver, we instead exploit the established link between adjoint-mode magnitude and local base-flow shear: prior studies of swirl-stabilized flows have shown that the adjoint mode is concentrated along the shear layer \cite{tammisola2016coherent}, and that the growth rate and wavemaker behavior of the global instability are governed primarily by the velocity gradient in the inner shear layer \cite{oberleithner2013nonuniform}. Motivated by this correspondence, we define a radial wavemaker indicator

\begin{equation}
    \mathcal{I}(r) =
        \left|\hat{u}(r)\right|_{\!\mathrm{norm}}
        \cdot 
            \,\left|\partial_r \bar{u}(y)\right|_{\!\mathrm{norm}}.
    \label{eq14}
\end{equation}

where the subscript ``norm'' denotes normalization by the respective field maximum, so that all contributions are bounded on $[0,1]$. The azimuthal shear contribution, $\partial_r \bar u_\theta$, was computed but not included in the final indicator, since the growth rate and wavemaker behavior of the global instability are governed primarily by the axial velocity gradient in the inner shear layer, with the azimuthal shear playing a comparatively minor role in setting the local receptivity of the flow; the indicator was therefore constructed using only the axial shear term, $\partial_r \bar u_y$.

The normalized eigenmode magnitude, $|\hat{u}(r)|_{\mathrm{norm}}$, represents the spatial distribution of the perturbation velocity amplitude obtained from the linear stability analysis, normalized by its maximum value. It characterizes the radial structure of the unstable global mode and identifies the regions where the instability-induced velocity perturbations attain their greatest amplitude. Higher values of $|\hat{u}(r)|_{\mathrm{norm}}$ indicate that the global mode exhibits a strong perturbation at the corresponding radial location. $|\partial_r \bar{u}(y)|_{\mathrm{norm}}$ is the radial gradient of the mean axial velocity, $\bar{u}(y)$, normalized by its maximum. Physically, this measures how steep the local shear is in the mean flow.  In a swirling flow, $|\partial_r \bar{u}(y)|_{\mathrm{norm}}$ peaks in the shear layers; specifically, the inner shear layer (between the central recirculation zone and the annular jet) and the outer shear layer (between the jet and the surrounding coflow/ambient). The non-dimensionalized radial wavemaker coordinate is reported as $r/D$. 

\subsection{Determination of Flame Attachment Location from OH$^*$-chemiluminescence Images}
\label{flame location}

OH$^*$-chemiluminescence images were acquired at a fixed operating condition over a sampling frequency of 2000 Hz. Each image was stored as a two-dimensional intensity array of size $n_x \times n_y$,
with the physical spatial coordinates $x$ (lateral) and $y$ (axial) provided alongside the intensity data. The time-averaged mean of OH$^*$-chemiluminescence intensity field was computed by time-averaging over at least $N_{OH} =$ 2000 frames:

\begin{equation}
    \bar{I}(x,y) = \frac{1}{N_{OH}}\sum_{k=1}^{N} I_k(x,y),
    \label{eq15}
\end{equation}

where $I_k(x,y)$ denotes the raw intensity of the $k$-th image. All subsequent analyses were performed on $\bar{I}(x,y)$.

To isolate regions of significant OH$^*$-chemiluminescence signal from the background, a global intensity threshold was applied to the mean intensity field \cite{bayley2012local}. The threshold level was defined as a fixed fraction of the global maximum intensity:

\begin{equation}
    I_{\mathrm{th}} = \eta \cdot \max_{x,y} 
    \ \bigl [\bar{I}(x,y)\bigr],
    \label{eq16}
\end{equation}

where $\eta = 0.20$ is the threshold ratio, selected to reliably delineate the flame-bearing regions from the non-reacting background. A binary flame mask was then constructed \cite{bayley2012local} as

\begin{equation}
    \mathcal{M}(x,y) =
    \begin{cases}
        1 & \text{if } \bar{I}(x,y) > I_{\mathrm{th}}, \\
        0 & \text{otherwise.}
    \end{cases}
    \label{eq17}
\end{equation}

The lateral position of the burner centerbody axis was identified directly from the spatial coordinate array $x = 0$. The centerline pixel index $x_c$ was located as the grid point at which the sign of $x$ changes from negative to positive, i.e.\ the zero-crossing of the lateral coordinate:

\begin{equation}
    x_c = \min\bigl\{j \;:\; x_j > 0\bigr\}.
    \label{eq18}
\end{equation}

This index was used to partition the image domain into a left branch ($x < 0$) and a right branch ($x > 0$) for the separate identification of the two flame attachment points.

The flame attachment location was determined independently for the left and right branches of the flame within a vertical search region spanning the lowermost $n_s = 100$ pixel rows above the burner nozzle exit plane. This search region was chosen to encompass the expected range of flame attachment heights while excluding regions far downstream where the flame is fully developed.

For each branch, the search was conducted by scanning upward row by row (increasing $y$) starting from the nozzle exit, and identifying the lowest axial position at which the flame mask $\mathcal{M}$ contains at least one active pixel in the respective lateral half of the domain. Specifically, at each axial row $y_j$ within $1 \leq j \leq n_s$, the last active pixel in the left half ($1 \leq x \leq x_c$) was identified. The first row for which such a pixel exists defines the left attachment point $(x_{\mathrm{att,L}},\,y_{\mathrm{att,L}})$. This procedure identifies the innermost active pixel closest to the centerbody axis on each side at the lowest detectable axial height, which physically corresponds to the flame root attachment location on the respective shear layer.

The lateral coordinates of the detected attachment points were non-dimensionalized by the hydraulic diameter of the burner nozzle, $D = 27.85$\,mm:

\begin{equation}
    x^*_{\mathrm{att,L}} = \frac{x_{\mathrm{att,L}}}{D},
    \qquad
    x^*_{\mathrm{att,R}} = \frac{x_{\mathrm{att,R}}}{D}.
    \label{eq19}
\end{equation}

The resulting non-dimensional attachment coordinates
$x^*_{\mathrm{att,L}}$ and $x^*_{\mathrm{att,R}}$ provide a measure of the lateral spread of the flame root relative to the nozzle diameter and facilitate direct comparison across operating conditions.

\section{Experimental setup}
\label{sec2}

\begin{figure}[!t]
\centering
\includegraphics[scale=0.222]{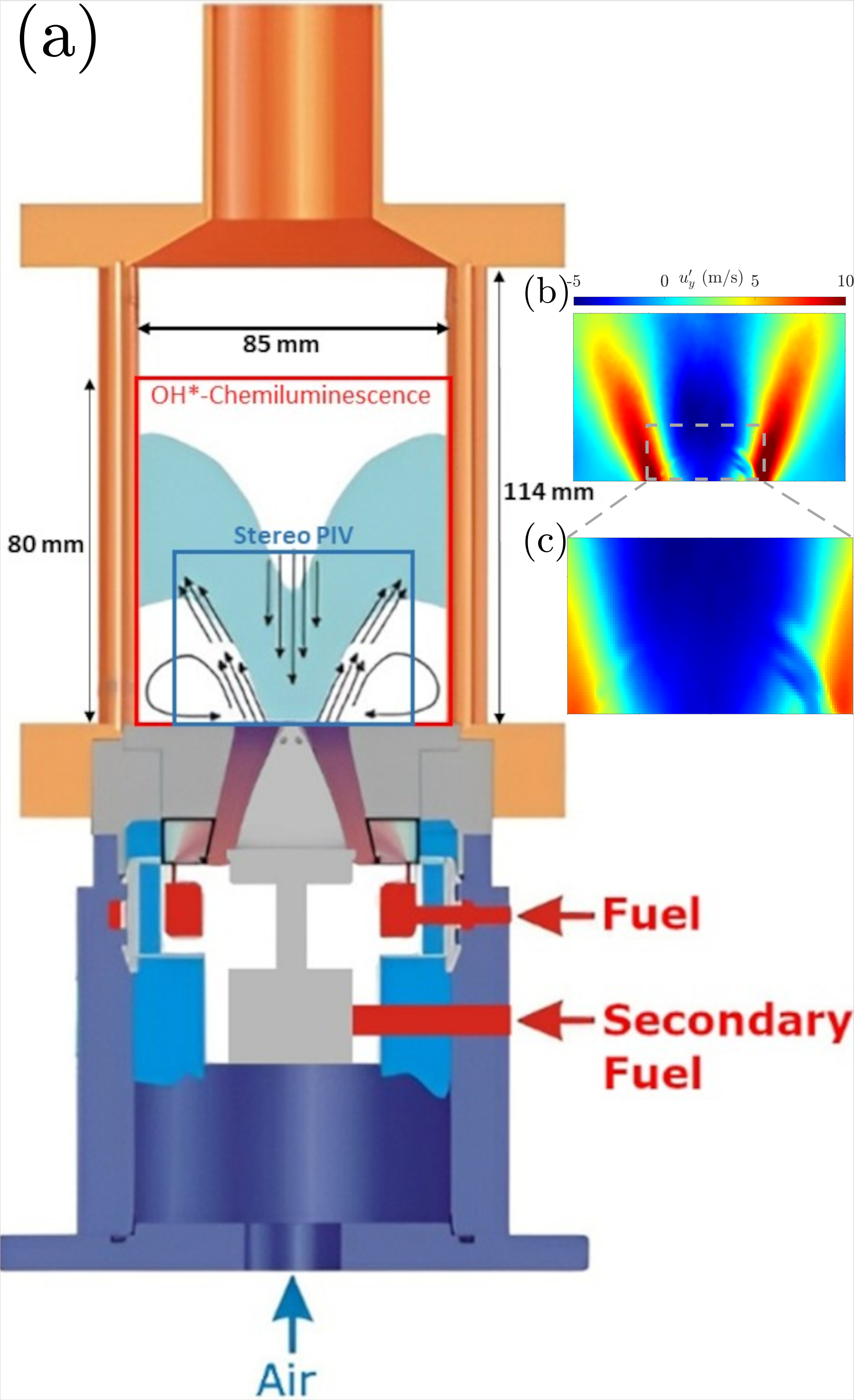}
\caption{ Schematic of the experimental setup (PRECCINSTA burner). (b) Mean non-reacting axial velocity field for the operating condition $\phi$ = 0.65 and $P_{th}$ = 10 kW. (c) Magnified view of the boxed region near the burner inlet, highlighting the reflection on the right side of the flow field. 
\label{fig01}}
\end{figure}

\subsection{Gas turbine model combustor}
\label{sec2.1}

Experimental studies have been carried out in a gas turbine model combustor, referred to as PRECCINSTA (see Fig. \ref{fig01}). PRECCINSTA stands for PREdiction and Control of Combustion INSTAbilities and is based on a real industry-based design from company Turbomeca and can be operated in either partially or fully premixed regimes. In the current experimental setup, the combustor operates under ambient pressure in partially premixed mode with a methane–air mixture as fuel. The reactants are initially fed into a plenum chamber and pass through a swirl generator that consists of 12 radial vanes. Methane is supplied perpendicularly to the air flow through holes situated in the swirler {channels, controlled with a Bronkhorst mass flow meter. Such a design corresponds to a jet-in-crossflow configuration. The swirling mixture flows into the combustor through a burner nozzle having a diameter of $D$ = 27.85 mm with a conical bluff body in its center. The combustor chamber is a square one with a size of 85 $\times$ 85 mm$^2$ and height of 114 mm. Quartz glass walls provide optical access and are fixed by metal posts in four corners. An outlet of the combustor chamber is a conical part followed by an exhaust duct of inner diameter 40 mm.

\subsection{Secondary injection and fuel-staging of methane}
\label{sec2.2}

Methane was supplied as a secondary fuel through six azimuthally equally spaced holes, each with a diameter of 2 mm, located on the circumference of the centerbody, positioned 5.31 mm upstream of the centerbody tip. The secondary fuel supply was managed by a mass flow meter operating under steady-state conditions, with the continuous flow initially flared through a separate, piloted afterburner. Immediately prior to image acquisition, a 3/2-way solenoid valve (Type 0330, Bürkert) was actuated to divert the fuel into the PRECCINSTA burner, where it was finally injected through the centerbody orifices.} The secondary injection velocity was maintained below 5\% of the bulk flow velocity to suppress direct momentum forcing while enabling controlled local modulation of the mixture fraction field in the vicinity of the reaction zone. This injection strategy was designed to alter the spatial distribution of heat release without significantly affecting the global flow topology.

Two distinct secondary-fuel injection approaches were employed. In the first approach, the secondary methane flow rate was progressively increased while maintaining the primary fuel flow rate constant. This resulted in an increase in both the total thermal power and the global equivalence ratio of the combustor. In the second approach, referred to as fuel-staging, a fraction of the methane supplied through the primary injector was redistributed to the secondary injection ports while keeping the total fuel flow rate constant (shown in Tab. \ref{tab1}). Consequently, the overall thermal power and global equivalence ratio remained unchanged, whereas the spatial distribution of fuel and heat release within the combustor was modified. These two approaches enabled independent investigation of the effects of increased heat input and fuel redistribution on flame dynamics. By systematically varying the secondary methane addition, the heat-release field could be manipulated, thereby enabling controlled variation of the spatial overlap between the flame and the wavemaker region under otherwise identical operating conditions.

\subsection{Stereoscopic particle image velocimetry}
\label{sec2.3}

Three-component velocity field measurements of the non-reacting flow were performed using stereoscopic particle image velocimetry (sPIV) at a frequency of 10 Hz for the combinations of different equivalence ratio and density (based on thermal power). The sPIV setup comprises a dual-pulse Nd:YAG laser (Quantel Evergreen EVG00200; energy per pulse is 150 mJ at 532 nm), two sCMOS cameras (Andor Zyla 5.5; each has 2560 px $\times$ 2160 pixels) and two pulse generators (Quantum model 9514 and Quantum model 9520 series). The tracer particles were produced using a medium-sized PIVlight30 particle generator and PIVLIGHT high quality seeding fluid (Dioctyl sebacate) from PIVTEC GmbH. The particles had a mean diameter of $d_p$ = 1.2 $\mu$m and have a response time of $\tau$ $\approx$ 4 $\times$ 10$^{-6}$ s. Considering the bulk flow velocity ($U$) to be 40 m/s, and the vortex diameter to be 20 mm, the Stokes number ($St$) = 0.008 $ \ll $ 0.1 ensures that the particles track the flow accurately. The laser beam was shaped into a light sheet covering the lower region of the combustion chamber using a pair of cylindrical lenses with focal lengths of -40 mm and 250 mm. The sheet was subsequently focused to a thin waist by a third cylindrical lens with a focal length of 1000 mm. The thickness of the resulting laser sheet was estimated to be approximately 1 mm. The lenses used on the cameras were an FF macro objective lens (f = 100 mm; f/5.6). The PIV measurement domain spans a region of 62$\times$ 35mm$^2$, -31 < $x$ < 31 mm and 0 < $y$ < 35 mm, shown in Fig. \ref{fig01}a.

For each of the operating conditions, the test was performed five times, leading to a total number of 400 velocity fields. The velocity fields were determined from the particle images by means of commercial PIV software (LaVision Davis 11.0). The multi-scale cross-correlation method was used with the interrogation window size of 16×16 pixels with an overlap of 50\%. Figure \ref{fig01}b shows the mean axial velocity field for $\phi$ = 0.65, $P_{th}$ = 10 kW, and Fig. \ref{fig01}(c) presents a magnified view of the boxed region near the burner inlet. A distinct diagonal artifact is visible in this region, attributed to a laser-light reflection off the burner surface near the burner inlet, which locally corrupts the PIV velocity signal. This reflection is confined to the right-hand side of the domain, while the left-hand side remains unaffected. Given this, only data collected on the left-hand side of the burner is used in subsequent analysis. 

\begin{table}[!t]
\caption{\label{tab1} Experimental operating conditions with secondary fuel injection rates and fuel-staging rates in standard liters per minute (SLPM) for different thermal powers and global equivalence ratios.}
\fontsize{8pt}{9pt}\selectfont
\begin{tabular}{|p{0.5 cm}|p{0.5 cm}|p{1 cm}|p{1.5 cm}|p{1 cm}|}
\hline
Serial & $\phi$ & $P_{th}$ (kW)  & Secondary Injection of fuel (SLPM) & fuel-staging (SLPM)\\
\hline
(a) &0.6  & 10  & 0 &  \\
\hline
  & &   & 0.25 &  \\
\hline
  & &    & 0.50 &  \\
\hline
  &  &   & 0.75 & 0.75 \\
\hline
  &  &   & 1.00 & 1.00 \\
\hline
  &  &   & 1.25 & 1.25 \\
\hline
(b)  &0.65  & 10  & 0 &  \\
\hline
  & &   & 0.25 & 0.25 \\
\hline
  & &    & 0.50 & 0.5 \\
\hline
  &  &   & 0.75 & 0.75 \\
\hline
  &  &   & 1.00 & 1.00 \\
\hline
(c)  &0.6  & 15  & 0 &  \\
\hline
  &   &   & 0.75 &  \\
\hline
  &  &   & 1.00 &  \\
\hline
& &   & 1.25 &  \\
\hline
  &  &   & 1.50 & 1.50 \\
\hline
  &  &   & 1.75 & 1.75 \\
\hline
  &  &   & 2.00 & 2.00 \\
  \hline
(d)  &0.65  & 15  & 0 &  \\
\hline
  & &   & 0.25 &  \\
\hline
  & &    & 0.50 &  \\
\hline
  &  &   & 0.75 & 0.75 \\
\hline
  &  &   &  & 1.00 \\
  \hline
  &  &   &  & 1.50 \\

\hline
\end{tabular}
\end{table}

\subsection{High-speed OH$^*$-Chemiluminescence imaging}
\label{sec2.4}

Line-of-sight integrated imaging of OH$^*$-chemiluminescence was utilized as a non-intrusive method for optical diagnosis to determine the spatial position of heat-release zone \cite{hardalupas2004local}. These observations were done through the use of an intensified CMOS camera (LaVision HSS8 with HS-IRO) that had a UV lens and UG-11 bandpass filter working in the spectral range of 300–325 nm. OH$^*$ radicals produced inside the reaction zone during the burning of fuel release ultra-violet light, therefore giving a nearly instantaneous depiction of flame position and heat-release zone intensity. Since OH$^*$ emission is very faint in the UV spectrum, the HS-IRO intensifier was operated with 70\% gain. An integration time of 200000 ns was used to enhance the signal-to-noise ratio (SNR). The experiments were performed at a frequency of 2 kHz, and depending on the conditions, a number of 6000-20000 images were taken. The heat release was computed using the integral summation of instantaneous images over a region of -40 $\leq x \leq $ 40 mm and 0 < $y \leq $ 80 mm. The spatial integral of this intensity over the imaging domain therefore serves as a surrogate measure of the instantaneous global heat-release rate $\dot{Q}(t) \approx \sum_{x,y} I(x,y,) $ \cite{lauer2010adequacy}. During combustion instability, the acoustic pressure signal oscillates in-phase with heat-release rate, reflecting the underlying thermoacoustic feedback loop \cite{nicoud2005thermoacoustic}.

\section{Results and Discussion}
\label{R&D}

\subsection{Operating conditions}
\label{Base OC}

We first present the flame dynamics observed in the combustor at two global equivalence ratios ($\phi$ = 0.60 and 0.65) and two thermal power levels ($P$ = 10 and 15 kW) \cite{kushwaha2021dynamical}. Subsequently, the influence of secondary methane injection and fuel-staging on the combustor dynamics is investigated. Particular emphasis is placed on how these fuel-injection strategies modify the flame behavior, heat-release distribution, and the resulting combustion dynamics of the system.

\begin{figure}[!t]
\centering
\includegraphics[scale=0.356]{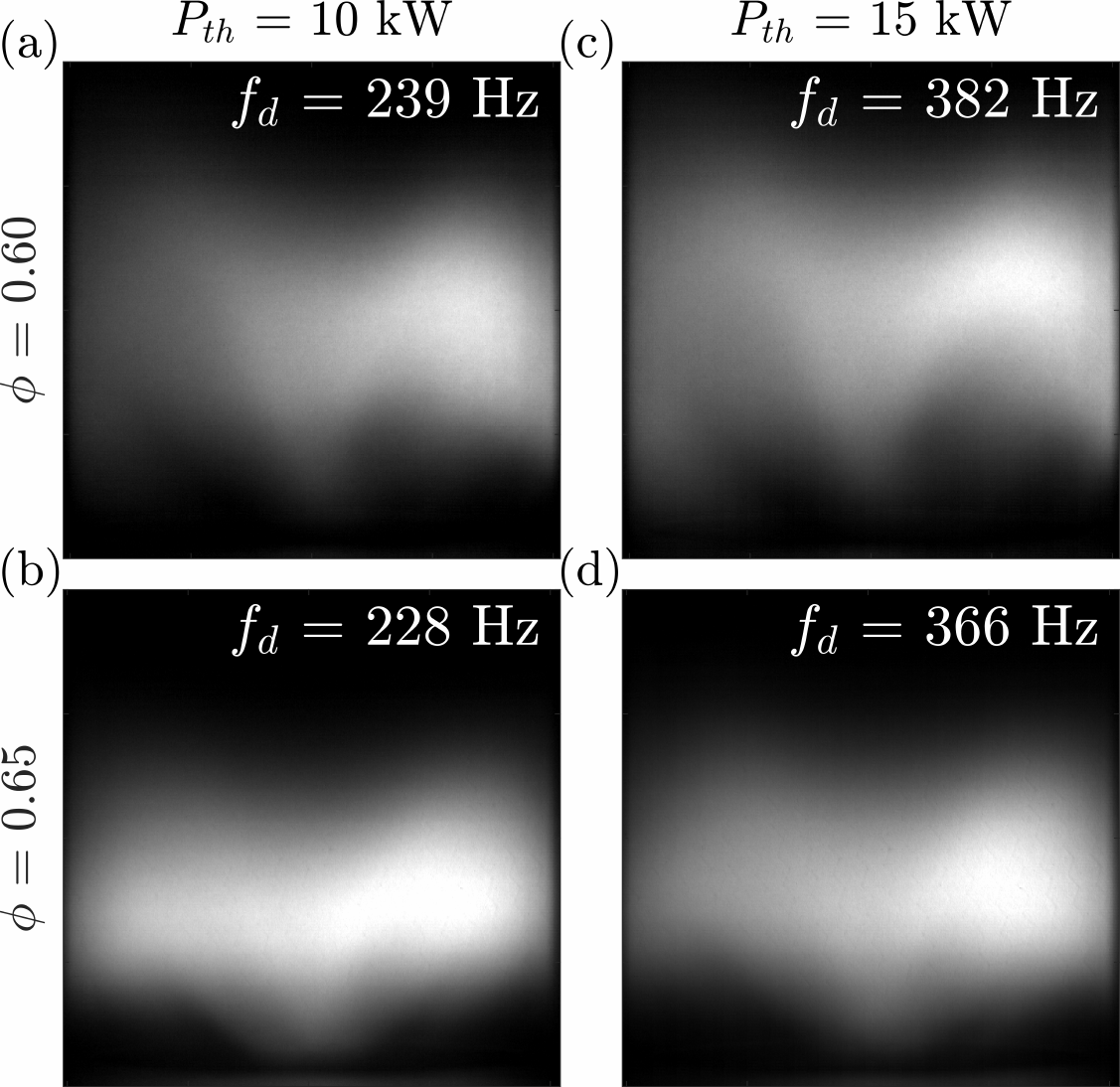}
\caption{Time-averaged OH$^*$-chemiluminescence images of the M-shaped (unstable combustion) flame under operating conditions:  (a) $\phi$ = 0.60, $P_{th}$ = 10 kW; (b) $\phi$ = 0.65, $P_{th}$ = 10 kW; (c) $\phi$ = 0.60, $P_{th}$ = 15 kW; (d) $\phi$ = 0.65, $P_{th}$ = 15 kW.
The dominant frequency $f_d$ is annotated for each case and is observed to increase with thermal power at fixed equivalence ratio.
\label{fig02}}
\end{figure}

\begin{figure*}[t!]
\centering
\includegraphics[scale=0.45]{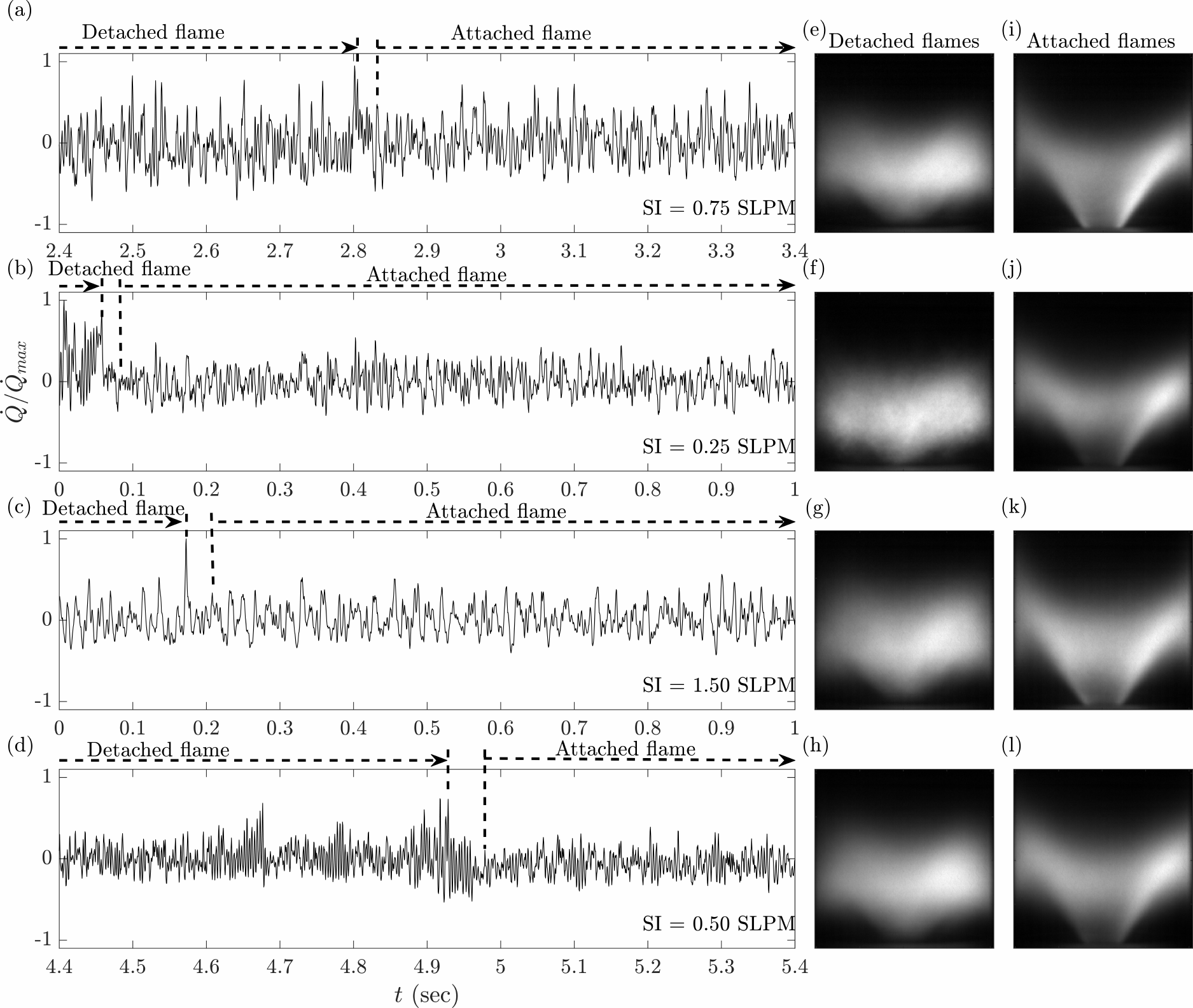}
\caption{Effect of secondary injection alone (without fuel-staging) on flame stabilization for the operating conditions (a) - (d), shown in Tab. \ref{tab1}.
Left Column:(a-d) The time series of the heat release signal, normalized by its maximum value; Middle Column:(e-h) the time-averaged OH$^*$-chemiluminescence images for the detached flame from the inlet and; Right Column:(i-l) for the attached flame to the inlet of the burner due to the secondary injection during the operating conditions, shown in Fig. \ref{fig02}. The amount of methane (in SLPM) used for secondary injection (SI) is provided in the bottom-right corner (left column) during the particular operating condition. The transition instants are marked by the vertical dashed line.
\label{fig03}}
\end{figure*}

Figure \ref{fig02} shows the behavior of the flame at the two equivalence ratios ($\phi$ = 0.60 and 0.65) and the two thermal power levels ($P$ = 10 and 15 kW), used as the operating conditions in the present study. These operating conditions are selected based on previous studies that indicate the presence of combustion instability. The time-averaged flame images in these operating conditions exhibit the M-shaped flames. The flames are detached from the inlet of the burner. Moreover, the flames are more lifted in case of $\phi$ = 0.60 in comparison to that in $\phi$ = 0.65. The dominating frequencies ($f_{d}$) during each operating condition are shown in the corresponding Fig. \ref{fig02}, is determined using the power spectrum of the heat release signals, as described in Section \ref{sec2.4}. The lower-thermal-power (10 kW) cases show dominant frequencies at 228 Hz and 239 Hz. However, the dominant frequencies shift to the higher values, i.e., 366 Hz and 382 Hz for the higher thermal power (15 kW), indicating that the dominant frequency is governed primarily by thermal power (flow rate) rather than equivalence ratio.

\subsection{Effect of secondary injection of fuel}
\label{secondary injection}

The effect of the secondary injection of fuel, i.e., methane, is depicted Fig. \ref{fig03} for the operating conditions (a) - (d), shown in Fig. \ref{fig02}. Secondary methane was injected at flow rates ranging from 0.25 to 2 SLPM, depending on the operating condition (see Tab. \ref{tab1}), corresponding to an increment in the global equivalence ratio, $\Delta \phi$, of 0.010–0.078. The cases shown in Fig. \ref{fig03} are the first cases which show the effect of the secondary injection. For the first three cases (a-c), the burner transitions to a stable V-shaped flame within 3 seconds of secondary injection (Fig. \ref{fig03}a-c). However, in the last case (d), with a thermal power of 15 kW and an equivalence ratio of 0.65, the burner transitions to stable operation at t > 4.9 sec (Fig. \ref{fig03}d). Prior to the transition, the heat-release rate exhibits large-amplitude, broadband fluctuations characteristic of the detached, M-shaped flame associated with combustion instability. Upon activation of secondary injection, the fluctuation amplitude drops markedly, coinciding with the transition to an attached, V-shaped flame. This behavior is consistent across all operating conditions, indicating that the onset of flame attachment is accompanied by a suppression of the heat-release oscillations that sustain the instability.

\begin{figure*}[!t]
\centering
\includegraphics[scale=0.452]{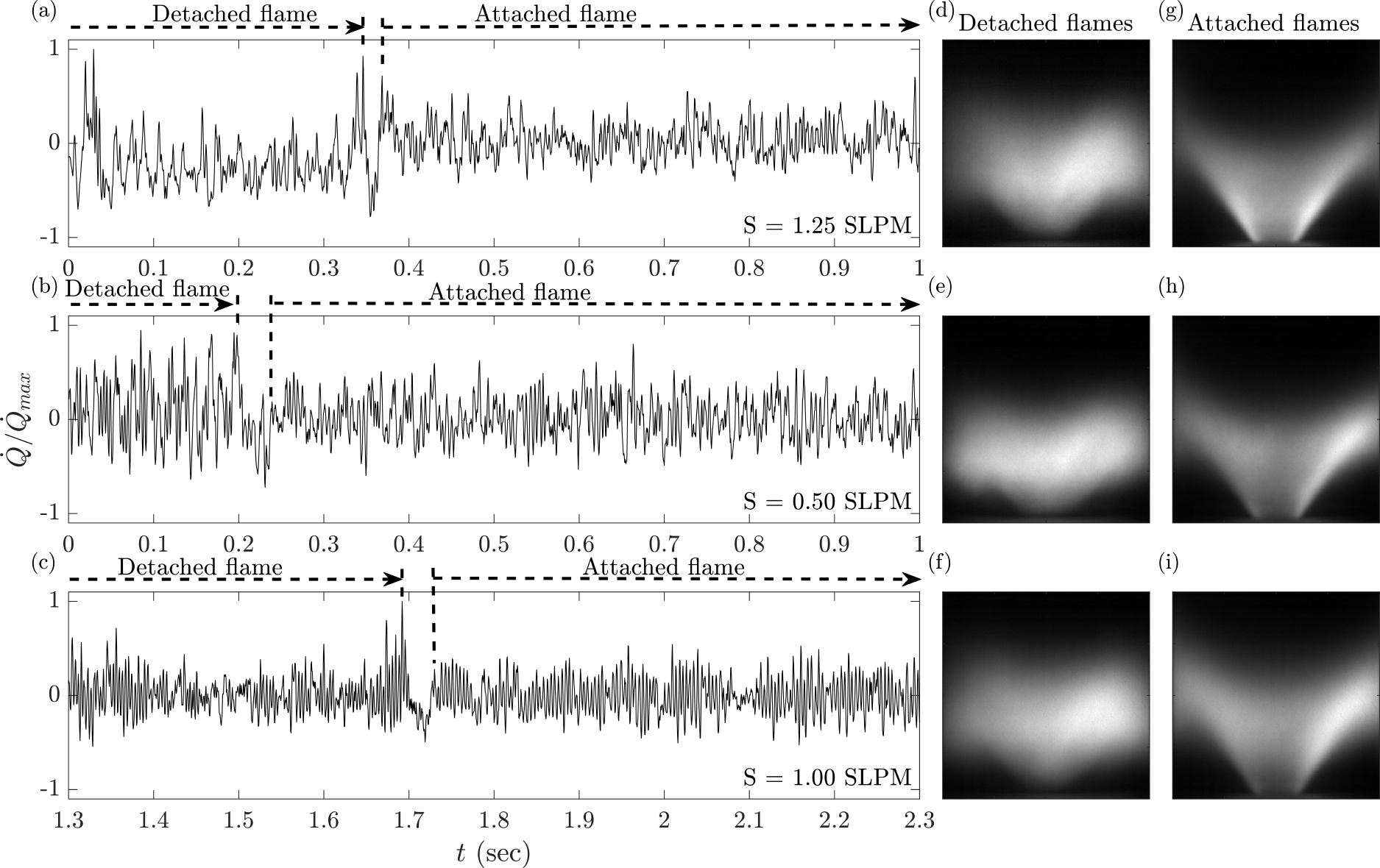}
\caption{Effect of fuel-staging on flame stabilization for three operating conditions: (a) $\phi$ = 0.60, $P_{th}$ =10 kW; (b) $\phi$ = 0.65, $P_{th}$ = 10 kW; (c) $\phi$ = 0.65, $P_{th}$ = 15 kW. Left Column:(a-c) The time series of the heat release signal, normalized by its maximum value, at the fuel-staging flow rate (S) annotated for each case. The transition instants are marked by the vertical dashed line. Middle Column:(d-f) the time-averaged OH$^*$-chemiluminescence images for the detached flame (unstable) from the inlet and; Right Column:(g-i) for the attached flame (stable) to the inlet of the burner due to the fuel-staging during the operating conditions $(a), (b)$ and $(d)$, shown in Fig. \ref{fig02}.
\label{fig04}}
\end{figure*}

The corresponding time-averaged OH$^*$-chemiluminescence images, shown for the detached-flame phase (Figs. \ref{fig03}e-h) and the attached-flame phase (Figs. \ref{fig03}i-l), confirm the associated M-to-V flame shape transition. In this configuration of secondary injection of methane, the total fuel flow rate, global equivalence ratio, and thermal power, all increase concurrently with the onset of secondary injection. The M-to-V flame transition observed here therefore reflects the combined effect of the additional heat release and its altered spatial distribution. The complementary fuel-staging, discussed in Section \ref{staging}, isolate the effect of heat-release redistribution separately by keeping the total fuel flow rate fixed constant.

\subsection{Effect of fuel-staging}
\label{staging}

Figure \ref{fig04} shows the effect of fuel-staging on flame stabilization for three operating conditions: (a) $\phi$ = 0.60, $P_{th}$ = 10 kW; (b) $\phi$ = 0.65,$P_{th}$ = 10 kW; and (c) $\phi$ = 0.65,$P_{th}$ = 15 kW. Unlike the secondary-injection-only cases, fuel-staging redistributes an equivalent amount of fuel from the primary to the secondary injection ports, holding the total fuel flow rate, global equivalence ratio, and thermal power fixed at their nominal operating values. The left column shows the time evolution of the global heat-release rate, normalized by its maximum value, with the required staging flow rate (S) annotated for each case: 1.25, 0.50, and 1.00 SLPM for (a)–(c), respectively. As in the secondary injection cases, the effect of fuel-staging occurs quickly (Figs. \ref{fig04}a-c) and the detached-flame exhibits large-amplitude, heat-release fluctuations, which drop sharply at the onset of flame attachment. The corresponding time-averaged OH$^*$-chemiluminescence images, shown for the detached-flame (Figs. \ref{fig04}d–f) and the attached-flame (Figs. \ref{fig04}g–i), confirm the same M-to-V flame shape transition observed with secondary injection alone.

The operating condition, characterized by $\phi$ = 0.6,$P_{th}$ = 15 kW (c, in Tab. \ref{tab1}), showed no discernible influence of the fuel-staging process even at a secondary fuel flow rate of 2 SLPM. Owing to safety constraints, further increases in the staging flow rate could not be pursued for this operating condition. Since the global equivalence ratio remained constant throughout, any change in flame shape at this condition cannot be attributed to a change in the overall mixture strength. Rather, flame stabilization via fuel-staging results from the redistribution of fuel within the combustor, which modifies the spatial distribution of the heat-release rate without altering the global fuel-air ratio.

Fuel-staging holds the total fuel flow rate constant, so these results isolate the effect of the spatial redistribution of heat release from any change in thermal power or global equivalence ratio. The observation that fuel-staging separately reproduces the same stabilizing M-to-V transition, seen with secondary injection (see Section  \ref{secondary injection}), for the three conditions where stabilization was achieved, demonstrates that flame stabilization is governed primarily by the relocation of heat release toward the wavemaker region, rather than by the concurrent increase in thermal power and equivalence ratio present in the secondary injection cases of Fig. \ref{fig03}.

\subsection{Flame–Wavemaker Spatial Overlap}
\label{wavemaker}

\begin{figure}[!t]
\centering
\includegraphics[scale=0.332]{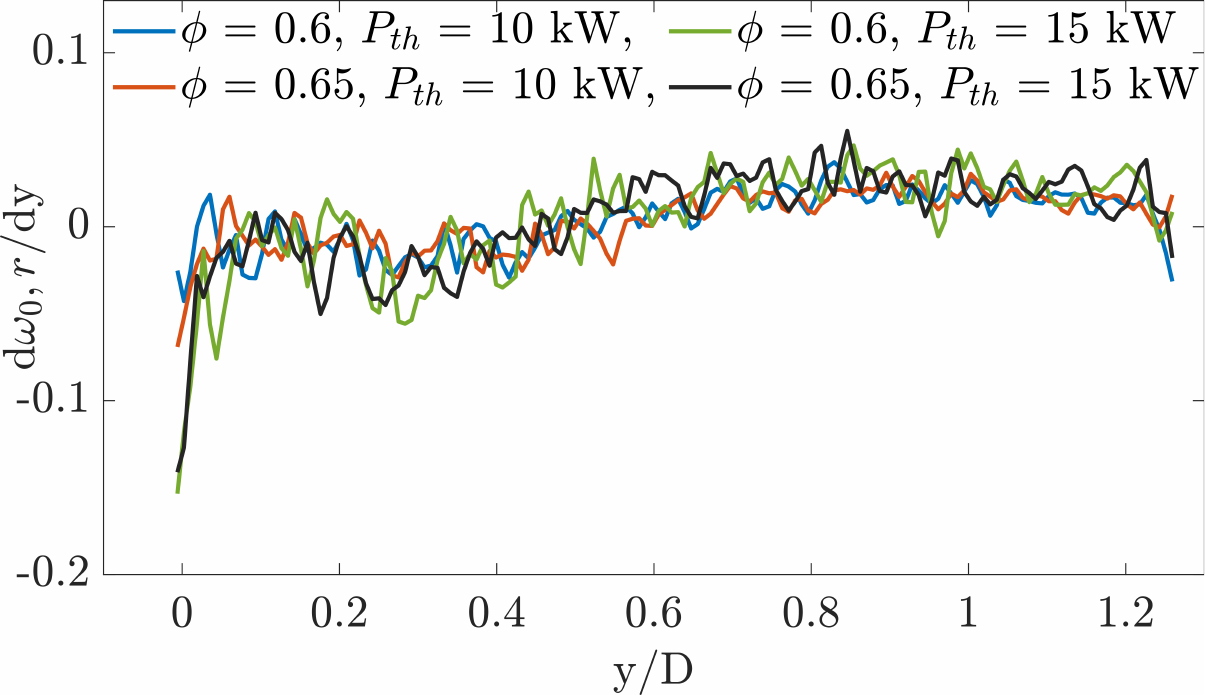}
\caption{Streamwise derivative of the absolute frequency, $|\mathrm{d}\omega_{0,r}/\mathrm{d}y|$, as a function of axial distance $y/D$ for the operating conditions (a) - (d). The zero-crossing, occurring at a consistent axial location near the combustor inlet for all cases, identifies the axial wavemaker station $y_s$ via the saddle-point criterion.
\label{fig05}}
\end{figure}

Figure \ref{fig05} shows the streamwise derivative of the absolute frequency, $\mathrm{d}\omega_{0,r}/\mathrm{d}y$, evaluated across the operating conditions, in accordance with the saddle-point criterion for the axial wavemaker location  \citep{oberleithner2015formation,oberleithner2013nonuniform}. For all cases, the derivative crosses zero at a consistent location near the combustor inlet, confirming that the global mode wavemaker is anchored at the inlet regardless of thermal power or equivalence ratio. This zero-crossing is used to define the axial wavemaker station $y_s$, and the radial location of the wavemaker region at this station is discussed later. The consistency of $y_s$ across all operating conditions confirms that the axial wavemaker location is largely insensitive to changes in density  (based on thermal power) and equivalence ratio.

It should be noted that the wavemaker location, i.e. $y/D \approx 0-0.06$, identified is derived from the non-reacting flow field, whereas the flame stabilization behavior discussed in Section \ref{flame location} occurs under reacting conditions. Heat release introduces density stratification that can, in principle, modify the local shear structure and the associated hydrodynamic stability characteristics of the flow \cite{oztarlik2020suppression}. Prior studies of swirl-stabilized flows have reported that the axial wavemaker location is comparatively robust to moderate density variation introduced by combustion  \cite{oberleithner2015formation, luckoff2019excitation}, with the reacting and non-reacting wavemaker positions remaining in close spatial agreement. On this basis, the non-reacting wavemaker is used here as a hydrodynamic reference frame against which the reacting-flow flame position is compared. This study does not directly establish that the reacting-flow wavemaker is identical to its non-reacting counterpart; rather, the central experimental finding is that the flame root consistently relocates into the same radial band independently identified from the non-reacting stability analysis, which we interpret as strong correlative evidence for the governing role of flame–wavemaker overlap.


\begin{figure*}[!t]
\centering
\includegraphics[scale=0.415]{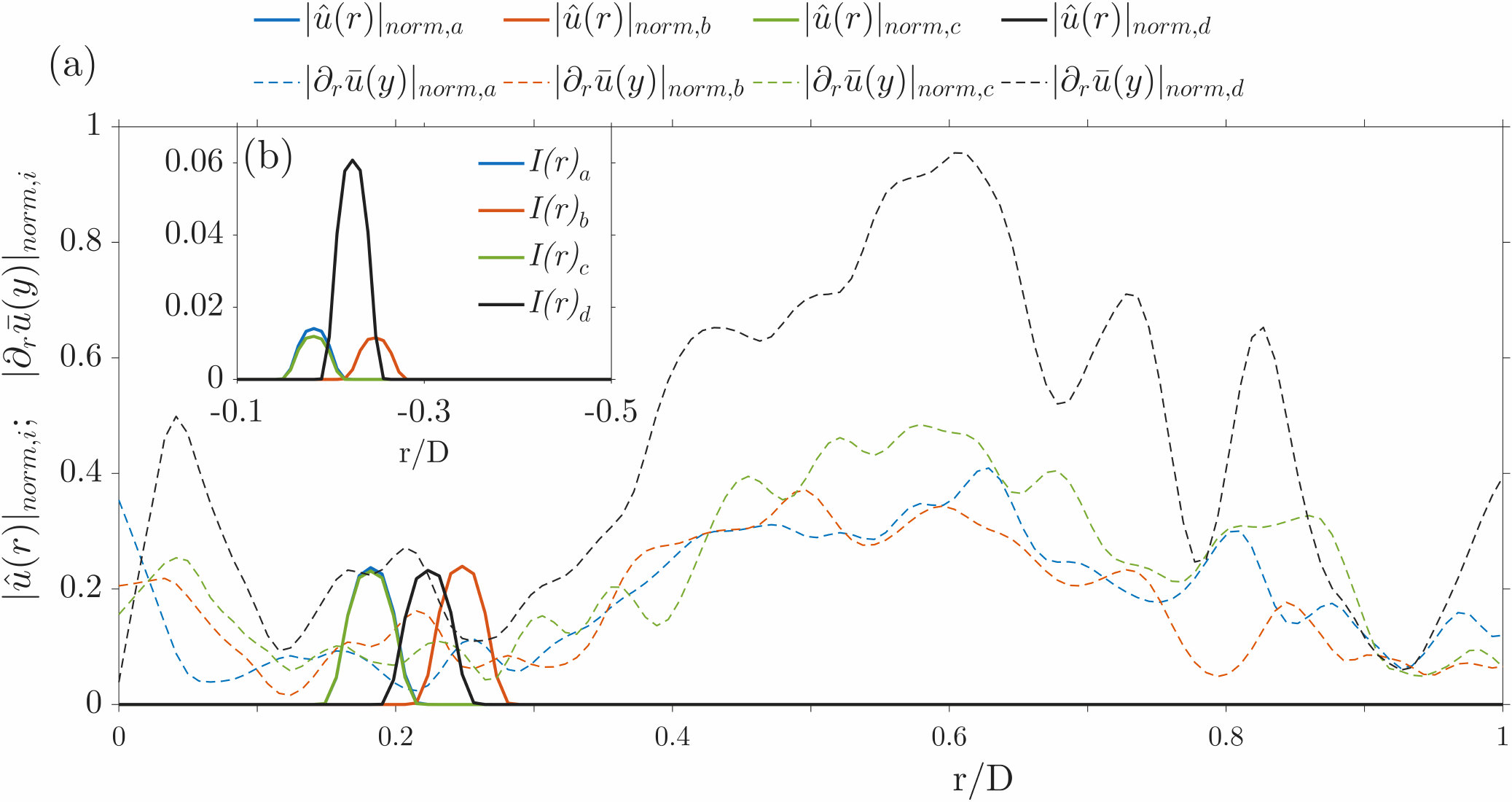}
\caption{(a) Variation of the instability mode amplitude $|\hat{u}(r)|_{\mathrm{norm}}$ (solid) and the radial gradient of the mean axial velocity ($|\partial_r \bar{u}(y)|_{\!\mathrm{norm}}$) (dashed), and (b) Variation of the wavemaker indicator $I(r)$, expressed in \ref{eq14}, over the non-dimensional radial direction (left half) for the operating conditions.
\label{fig06}}
\end{figure*}

Figure \ref{fig06}a compares the two components of the wavemaker indicator, $|\hat{u}(r)|_{\mathrm{norm}}$ and $|\partial_r \bar{u}(y)|_{\!\mathrm{norm}}$, across the representative cases (a–d). The instability mode amplitude $|\hat{u}(r)|_{\mathrm{norm}}$ is sharply localized within a narrow radial band (-0.30 $\leq r/D \leq$ -0.15) for all cases, whereas the base-flow shear $|\partial_r \bar{u}(y)|_{\!\mathrm{norm}}$ is broadly distributed and exhibits multiple local maxima across the domain, reflecting the presence of several shear layers in the confined swirling flow. Because the shear term alone does not uniquely identify a single radial location, the product $\mathcal{I}(r)$ inherits its spatial selectivity primarily from the mode amplitude, which effectively acts as a spatial window restricting the indicator to the region where the instability mode itself has non-negligible support. As shown in Fig. \ref{fig06}b, this results in a consistently narrow, single-peaked $\mathcal{I}(r)$ across all cases, confirming that the wavemaker location identified by the product is robust and is not an artifact of the broader, multi-peaked shear field alone.

 The wavemaker indicator $\mathcal{I}(r)$ peaks at $r/D$ = -0.182 for both (a) and (c), at $r/D$ = -0.248 for (b), and at $r/D$ = -0.223 for (d) (in Fig. \ref{fig06}b). Notably, the peak location is identical between (a) and (c), cases sharing the same equivalence ratio ($\phi$ = 0.6) but different densities (based on thermal power), indicating that the radial wavemaker position is primarily governed by $\phi$ rather than by thermal power. In contrast, increasing $\phi$ from 0.6 to 0.65 shifts the peak radially outward, from $r/D$ = -0.182 to -0.248 at $P_{th}$ = 10 kW (a→b), and from $r/D$ = -0.182 to -0.223 at $P_{th}$ = 15 kW (c→d). This consistent outward shift with increasing equivalence ratio suggests that richer mixtures displace the wavemaker further from the centerbody, plausibly reflecting a broader or radially displaced shear layer at higher equivalence ratio. Despite these shifts, all peaks remain within the (-0.30 $\leq r/D \leq$ -0.15) band identified as the wavemaker region from the stability analysis, and case (d) additionally exhibits the sharpest, most narrowly localized $\mathcal{I}(r)$ peak (inset (b)), consistent with the elevated base-flow shear observed across the radial domain at the highest thermal power and equivalence ratio.



\begin{figure}[!t]
\centering
\includegraphics[scale=0.256]{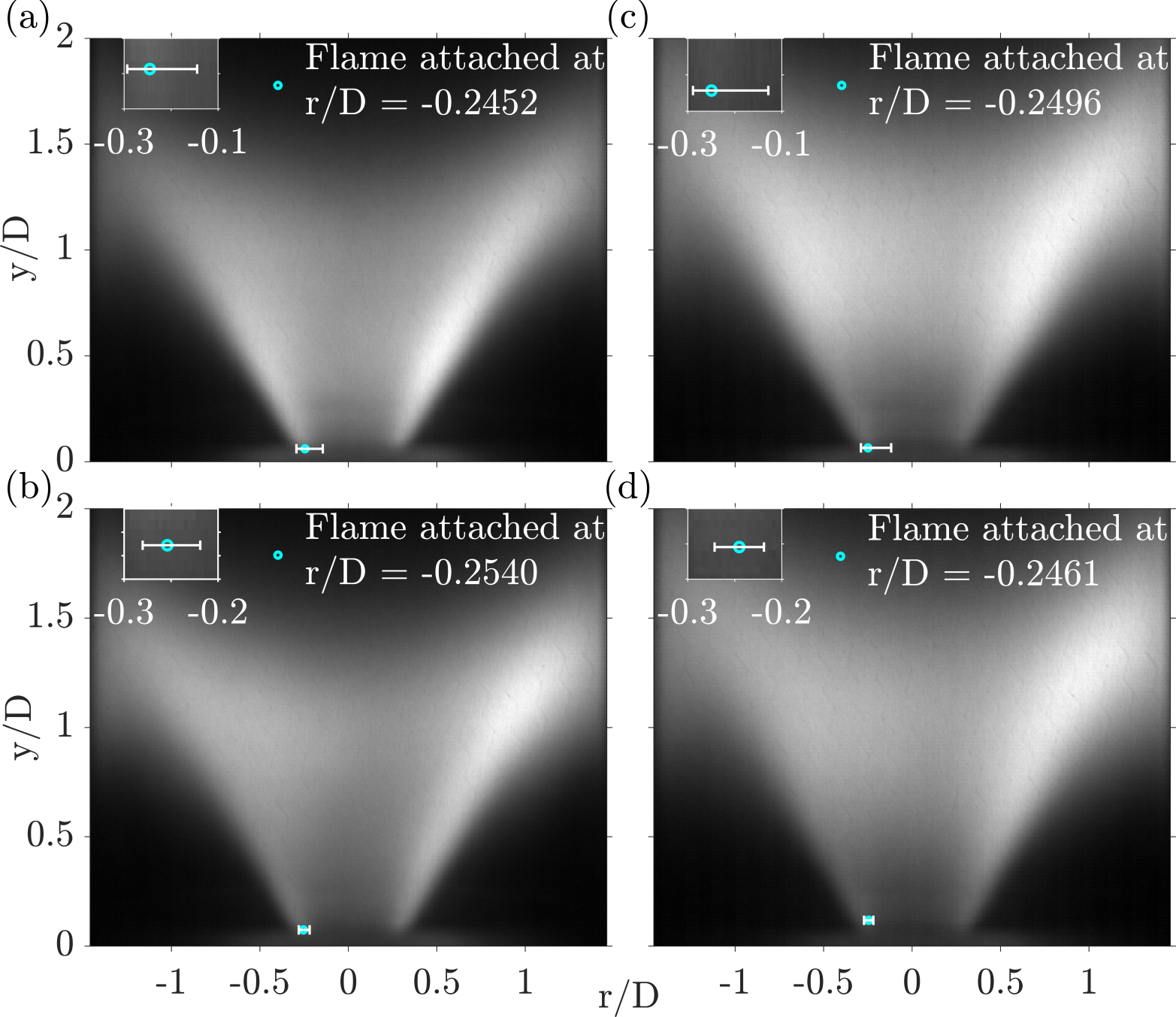}
\caption{Time-averaged, OH$^*$-chemiluminescence images for the secondary-injection and fuel-staging during the operating conditions (a)–(d), with the extracted flame attachment location (cyan circle) and its radial range (white bar) annotated for each case. Insets show zoomed views of the attachment region. 
\label{fig07}}
\end{figure}

Figure \ref{fig07} shows the time-averaged OH$^*$-chemiluminescence images for the secondary-injection and fuel-staging of the methane during the operating conditions (a)–(d), together with the extracted flame attachment location (cyan circle) and its associated radial range (white bar), with the inset providing a zoomed view of the attachment region. Since the OH$^*$-chemiluminescence intensity is proportional to the local heat-release rate, the attachment points correspond to the upstream limit of the heat-release zone. The extracted attachment locations are $r/D$ = -0.2452, -0.2540, -0.2496, and -0.2461 for cases (a)–(d), respectively, clustering tightly within a narrow band of $r/D$ $\approx$ -0.245 to -0.254 across all the operating conditions. These locations are superimposed on the wavemaker regions extracted for the corresponding operating conditions from the non-reacting velocity data across the range of secondary injection and fuel-staging of methane, and coincide closely with the radial wavemaker band identified independently from the linear stability analysis (-0.30 $\leq r/D \leq$ -0.15).

Tracked as a function of increasing secondary injection, the flame attachment position is found to migrate progressively into this radial band, with stable flame configurations consistently anchoring within the bounds of the wavemaker region. Conversely, during combustion instability, the flame remains lifted, corresponding to the M-shaped flame topology, and exhibits greater sensitivity to flow perturbations. The close correspondence between the radial extent of $\mathcal{I}(r)$ and the flame attachment locations, obtained independently from reacting-flow imaging, provides direct experimental confirmation that secondary injection relocates the flame root into the wavemaker region, and that the wavemaker identified from the non-reacting base flow remains a robust predictor of flame stabilization behavior under reacting conditions.

Since the OH$^*$-chemiluminescence intensity is proportional to the local heat-release rate, the extracted attachment points mark the upstream limit of the heat-release zone. Fig. \ref{fig07} superimposes these locations on the wavemaker regions identified for operating conditions ($a-d$) across the range of secondary injection and fuel-staging of methane. The independent, reacting-flow measurement reproduces the trend established above and attachment points cluster within the wavemaker band under stable operation and depart from it under the unstable, M-shaped topology. A separate diagnostic OH$^*$-chemiluminescence imaging rather than sPIV/LSA recovers the same radial correspondence, reinforces that the non-reacting wavemaker is a robust predictor of flame-stabilization behavior independent of the measurement technique used to locate the flame under reacting conditions.

\section{Conclusions}

The present study experimentally investigated the influence of low-momentum secondary methane injection on flame stabilization in a swirl-stabilized combustor and examined the hypothesis that the spatial overlap between the flame and the wavemaker region is a necessary prerequisite for combustion stability in flows characterized by a wavemaker region. To isolate the effect of flame positioning from changes in the global operating conditions, methane was injected through circumferential holes in the centerbody at a velocity less than 5\% of the bulk flow velocity, while maintaining the overall thermal power through appropriate fuel-staging.

The experimental results demonstrated that secondary injection plays an important role in altering the flame structure as the flame transitions from M-type to V-type, leading to stable combustion. A similar transition was observed during fuel-staging, confirming that the changes arise primarily from the redistribution of heat release rather than from an increase in the total fuel flow rate or thermal power.

Linear stability analysis of the non-reacting flow fields showed that the wavemaker region remains anchored near the inlet of the combustion chamber. With the introduction of the secondary injection and fuel-staging of methane, the flame undergoes a substantial displacement, with the flame attachment shifting toward the wavemaker region. After the stable operation, the flame root consistently occupies the radial position of approximately (-0.30 $\leq r/D \leq$ -0.15), coinciding with the radial extent of the wavemaker identified through the stability analysis. These observations indicate that the stabilization mechanism is governed primarily by the relative spatial arrangement of the flame and the hydrodynamic instability source rather than by a displacement of the wavemaker itself.


The findings provide strong experimental support for the proposed hypothesis that the spatial overlap between the flame and the wavemaker region constitutes a necessary condition for stable combustion in swirl-stabilized flows exhibiting a wavemaker region. These results also suggest that secondary fuel injection functions as an effective passive-control strategy by relocating the flame toward the hydrodynamically sensitive region without significantly altering the underlying flow instability characteristics. Consequently, combustion stability can be suppressed and mitigated through targeted manipulation of the flame position rather than by modifying the global flow field.

\section*{Credit authorship contribution statement}

\textbf{Abhishek Kushwaha}: Writing - original draft, methodology, Data curation, Investigation, Funding acquisition. \textbf{Bjarne Lieth}: Data curation
\textbf{Isaac Boxx}: Review, Supervision.

\section*{Declaration of competing interest}

The authors declare that they have no known competing financial interests or personal relationships that could have appeared to influence the work reported in this paper.

\section*{Acknowledgments}

We thank Dr.-ing. Christoph Steinhausen for his administrative contribution in the project. The authors also thank Mr. Edgar Brauers, Mr. Thomas Bungert and Mr. Bernhard Müller for their help in finalizing the experimental setup. This project has received funding from the Deutsche Forschungsgemeinschaft (DFG) under the Walter Benjamin Fellowship 2024 (Project No. 541611060). 

\section*{Declaration of generative AI and AI-assisted technologies in the manuscript preparation process}

During the preparation of this work the author(s) used CHAT-GPT in order to improve structure and language of the text. After using this tool/service, the author(s) reviewed and edited the content as needed and take(s) full responsibility for the content of the published article.

\FloatBarrier

\bibliographystyle{cnf-num}
\bibliography{cnf-refs}

\end{document}